\documentclass[prd,twocolumn,superscriptaddress,altaffilletter,showpacs,nofootinbib]{revtex4}
\usepackage{amssymb}
\usepackage[dvips]{graphicx}
\usepackage{amsmath}

\newcommand{\be}{\begin{equation}}
\newcommand{\ee}{\end{equation}}
\newcommand{\bea}{\begin{eqnarray}}
\newcommand{\eea}{\end{eqnarray}}
\newcommand{\bib}{\bibitem}
\newcommand{\der}{\partial}
\newcommand{\vphi}{\varphi}

\begin{document}

\title{Scale invariance: fake appearances}

\author{Israel Quiros}\email{iquiros6403@gmail.com}\affiliation{Departamento de Ingenier\'ia Civil, Divisi\'on de Ingenier\'ia, Universidad de Guanajuato, Gto., M\'exico.}

\date{\today}

\begin{abstract}
In this paper we shall show that, unless the affine geometrical structure of the underlying spacetime manifold is specified, there is an ambiguity in the understanding of the scale invariance -- also Weyl invariance -- of the given theory of gravity. In this regard we cast doubts on several well-known theories which are claimed to be scale-invariant setups. We shall show that in a consistent scale invariant theory not only the action but also the geometrical laws which govern the affine structure of the underlying spacetimes are invariant under the Weyl gauge transformations.
\end{abstract}


\maketitle

\section{introduction}\label{intro}

Scale invariance -- also Weyl gauge invariance -- is being, once again, the subject of intensive debate \cite{bars, bars', bars'', kallosh-linde, carrasco-kallosh, bars-ult, padilla, percacci, prester, shaposhnikov, scholz, indios, quiros-2000, quiros-2013, quiros-2014}. It was Weyl \cite{weyl} who made the first serious attempt to create a scale-invariant theory of gravity (and of electromagnetism). Due to an unobserved broadening of the atomic spectral lines this attempt had a very short history \cite{perlick, novello, scholz-h}. Resurrection of scale invariance is associated with the following prototype action \cite{deser} (see also the latter \cite{smolin}):

\bea S=\int d^4x\sqrt{|g|}\left[\frac{\phi^2}{12}\,R+\frac{1}{2}(\der\phi)^2+\frac{\lambda}{12}\,\phi^4\right],\label{deser-action}\eea where $\phi$ is the dilaton\footnote{The role of the dilaton $\phi$ is to modify the local strength of gravity through the effective gravitational coupling $G(\phi)=(3/4\pi)\phi^{-2}$.} and $(\der\phi)^2\equiv g^{\mu\nu}\der_\mu\phi\der_\nu\phi$. Since, under the Weyl gauge transformations\footnote{In this paper we shall use interchangeably the terms Weyl gauge transformations and (local) scale transformations.}

\bea g_{\mu\nu}\rightarrow\Omega^{-2}g_{\mu\nu},\;\phi\rightarrow\Omega\,\phi,\label{scale-t}\eea where the non-vanishing smooth function $\Omega^2=\Omega^2(x)$ is the conformal factor, the combination $\sqrt{|g|}[\phi^2R+6(\der\phi)^2]$ is kept unchanged -- as well as the scalar density $\sqrt{|g|}\phi^4$ -- then the action (\ref{deser-action}) is invariant under (\ref{scale-t}). Any scalar field which appears in the gravitational action the way $\phi$ does, is said to be conformally coupled to gravity. Hence, for instance, the following action \cite{bars, bars', bars'', bars-ult, prester}:

\bea S=\int d^4x\sqrt{|g|}\left[\frac{\left(\phi^2-\sigma^2\right)}{12}\,R+\frac{1}{2}(\der\phi)^2-\frac{1}{2}(\der\sigma)^2\right],\label{bars-action}\eea is also invariant under (\ref{scale-t}) since both $\phi$ and $\sigma$ are conformally coupled to gravity, provided that the additional scalar field $\sigma$ transforms in the same way as $\phi$: $\sigma\rightarrow\Omega\,\sigma$. For the coupling $\propto (\phi^2-\sigma^2)^{-1}$ to be positive and the theory Weyl-invariant, the scalar $\vphi$ must have a wrong sign kinetic energy -- just like in (\ref{deser-action}) -- potentially making it a ghost. However, the local Weyl gauge symmetry compensates, thus ensuring the theory is unitary \cite{bars, bars', bars''}.

In general any theory of gravity can be made Weyl-invariant by introducing a dilaton. In \cite{percacci} it is shown how to construct renormalization group equations for such kind of theories, while in \cite{odintsov} it has been shown that scale invariance is very much related with the effect of asymptotic conformal invariance, where quantum field theory predicts that theory becomes effectively conformal invariant. In Ref. \cite{padilla} the authors present the most general actions of a single scalar field and two scalar fields coupled to gravity, consistent with second order field equations in four dimensions (4D), possessing local scale invariance.\footnote{I want to underline that, whenever I cite \cite{padilla} in the present work as an example of papers where the geometrical aspect of the gravitational theories is not explored in due details, it should be recognized that the aim of the authors of that paper was to construct scale invariant actions without making emphasis in the geometrical aspect of the corresponding theories.} It has been shown that Weyl-invariant dilaton gravity provides a description of black holes without classical spacetime singularities \cite{prester}. Singularities appear due to ill-behavior of gauge fixing conditions, one example being the gauge in which theory is classically equivalent to standard General Relativity (GR). In \cite{bars} (see also \cite{bars', bars''}) the authors show how to lift a generic non-scale invariant action in Einstein frame into a Weyl-invariant theory and present a new general form for Lagrangians consistent with Weyl symmetry. Advantages of such a conformally invariant formulation of particle physics and gravity -- claim the authors -- include the possibility of constructing geodesically complete cosmologies. In this regard see critical comments in \cite{carrasco-kallosh} and the reply \cite{bars-ult}.

A noticeable deficiency of several works on scale invariance -- among them those of references \cite{bars, bars', bars'', kallosh-linde, carrasco-kallosh, bars-ult, padilla, percacci, prester, shaposhnikov, deser} -- is the lack of any kind of discussion about whether the Weyl symmetry of the action is shared by the geometrical laws which govern the affine structure of the underlying manifold. That this poses an actual problem for theories claimed to be Weyl-invariant will be evident from the discussion in this paper. Recall that any scalar-tensor theory -- this includes the effective string theory \cite{wands} -- being a metric theory of gravity, is deeply connected with the geometric structure of the spacetime manifold. Besides the lack of discussion on the geometrical structure of given theories, in these works it is not discussed how the matter degrees of freedom, other than the conformally coupled scalars and radiation, impact on the scale invariance of a given theory. As we shall show this is a critical issue when discussing on Weyl invariance.

Aim of the present work is to fill the existing gap in the understanding of scale invariance by focusing in geometrical as well as in physical aspects of the issue. It will be shown, in particular, that several theories claimed to be Weyl-invariant do not admit matter fields other than radiation to couple minimally to the dilaton. Only a convenient non-minimal coupling of matter is allowed, which means that these matter degrees of freedom necessarily suffer the effects of an additional five-force (5-force) of non-gravitational origin thus destroying any existing Weyl invariance. As a matter of fact the theories whose action is Weyl gauge invariant but which are associated with Riemannian spacetime manifolds can not be actually scale-invariant since the affine properties of the Riemann space are modified by the scale transformations (\ref{scale-t}). This is reflected in that the equations of motion of time-like test particles, as well as the equations of motion of matter fluxes (continuity equations), are not scale-invariant. In this connection we shall show that, as anticipated by Dicke \cite{dicke}, generalizations of Riemann geometry -- notably Weyl(integrable) geometry \cite{quiros-2000, quiros-2013, quiros-2014, cheng, bib-weyl, books} -- seem to be a better suited arena where to play the scale-invariant laws of gravity. 

To start with in the next section we shall explicitly show the very well known -- yet usually forgotten -- fact that the gravitational action (\ref{deser-action}) (the same for (\ref{bars-action})) is nothing but plain Brans-Dicke (BD) theory \cite{bd} with the anomalous value $\omega=-3/2$ of the BD coupling. As it is also well-known, this means that only traceless matter (radiation) can be consistently treated in this theory. Other matter degrees of freedom can not be included in the action in a consistent way unless an appropriate non-minimal coupling with the dilaton is allowed. In sections \ref{riem} and \ref{wing} we explore the geometrical aspect of the Weyl gauge symmetry, an issue which is usually avoided when scale invariance is investigated. In section \ref{riem} we shall show that even the vacuum theory (\ref{deser-action}) (also (\ref{bars-action})) is not actually a scale-invariant setup. This is due to the fact that the equation of motion of time-like point-particles (properly the geodesic curves of the Riemann geometry) are not Weyl-invariant. Other aspects of scale-invariance such as the arising gauge freedom and the possibility of avoidance of several spacetime singularities are also discussed in this section. In section \ref{wing} we investigate the impact of a slight modification of Riemann geometry known as Weyl-integrable geometry (WING) in the study of scale invariance. Given that the laws of Weyl geometry, including the affine properties of space, are invariant under the scale transformations (\ref{scale-t}), it is not surprising that WING can be a natural arena where to investigate the Weyl invariance of the gravitational laws. It will be shown that, if associate a given scale-invariant gravitational action with WING spacetime manifolds, the resulting theory of gravity is actually Weyl-invariant. This result is independent of the matter content of the spacetime. The very debated issue of the possibility to avoid several spacetime singularities just by going into a different gauge will be discussed in section \ref{sing} by means of the study of several gauge-free curvature invariants. Physical discussion of the results and brief conclusions will be given in section \ref{conclu}.

In this paper, for sake of simplicity, we shall focus in theories whose gravitational sector is depicted by the prototype action (\ref{deser-action}). It is evident that the obtained results are safely applicable to any theory where the dilaton and perhaps other scalar fields are conformally coupled to the curvature as, for instance, in (\ref{bars-action}) (see the references \cite{bars, bars', bars'', bars-ult, padilla, prester}). Our analysis is fully classical so that any possible influence of quantum aspects on scale invariance is not considered here.

\section{Anomalous Brans-Dicke coupling}\label{anomal}

A first step towards a precise understanding of the meaning of scale invariance in theories where the gravitational sector is described by the action (\ref{deser-action}) is to realize that the above theory is nothing but plain BD theory \cite{bd} with the anomalous value of the BD coupling parameter $\omega=-3/2$. Actually, under the replacement $$\vphi\rightarrow\frac{\phi^2}{12}\;\Rightarrow\;\sqrt\frac{3}{\vphi}\,\der_\mu\vphi\rightarrow\der_\mu\phi,$$ the action (\ref{deser-action}) is written as (here we omit the quartic potential term) $$S\rightarrow S_\text{BD}=\int d^4x\sqrt{|g|}\left[\vphi\,R+\frac{3}{2}\frac{(\der\vphi)^2}{\vphi}\right].$$ 

To understand what this entails, let us consider the action (\ref{deser-action}) with the addition of the following matter piece:

\bea S_\text{mat}=\int d^4x\sqrt{|g|}\,{\cal L}_\text{mat}[\chi,\nabla\chi,g],\label{matter-action}\eea where ${\cal L}_\text{mat}$ is the matter Lagrangian and $\chi$ collectively stands for the matter degrees of freedom other than the conformally coupled scalars.\footnote{It is assumed that under (\ref{scale-t}) (see, for instance, the appendix A of Ref. \cite{wands}) $${\cal L}_\text{mat}[\chi,\nabla\chi,g]\rightarrow\Omega^4{\cal L}_\text{mat}[\chi,\nabla\chi,\Omega^{-2}g],$$ where it is explicit that $\sqrt{|g|}\,{\cal L}_\text{mat}$ is unchanged by the Weyl gauge transformations (\ref{scale-t}). Yet, implicitly it is seen that the particles couple to the conformal metric, which results in that these do not follow geodesics (see the discussion below).} The field equations which can be derived from the resulting total action

\bea S_\text{tot}=\int d^4x\sqrt{|g|}\left[\frac{\phi^2}{12}\,R+\frac{1}{2}(\der\phi)^2+\frac{\lambda}{12}\,\phi^4+{\cal L}_\text{mat}\right],\label{tot-action}\eea by taking variations with respect to $g_{\mu\nu}$ and $\phi$ which vanish on the boundary, are the following Einstein equations plus the Klein-Gordon (KG) equation for the dilaton $\phi$ respectively [$G_{\mu\nu}\equiv R_{\mu\nu}-g_{\mu\nu}R/2$]:

\bea &&G_{\mu\nu}=\frac{6}{\phi^2}\,T^\text{mat}_{\mu\nu}-\frac{4}{\phi^2}\left[\der_\mu\phi\der_\nu\phi-\frac{1}{4}g_{\mu\nu}(\der\phi)^2\right]\nonumber\\
&&\;\;\;\;\;\;\;\;\;\;\;\;\;\;+\frac{2}{\phi}\left[\nabla_\mu\nabla_\nu\phi-g_{\mu\nu}\Box\phi\right]+\frac{\lambda}{2}g_{\mu\nu}\phi^2,\label{feq}\\
&&\Box\phi-\frac{1}{6}\,R\phi=\frac{\lambda}{3}\,\phi^3,\label{kg-eq}\eea where $\Box\phi\equiv g^{\mu\nu}\nabla_\mu\nabla_\nu\phi$, and $$T^\text{mat}_{\mu\nu}=-\frac{2}{\sqrt{|g|}}\frac{\der\left(\sqrt{|g|}{\cal L}_\text{mat}\right)}{\der g^{\mu\nu}},$$ is the stress-energy tensor of matter. The trace of (\ref{feq}) $$\Box\phi-\frac{1}{6}\,R\phi=\frac{1}{\phi}\,T^\text{mat}+\frac{\lambda}{3}\,\phi^3,$$ when compared with the KG equation (\ref{kg-eq}) yields to the constraint $T^\text{mat}\equiv g^{\mu\nu}T^\text{mat}_{\mu\nu}=0$. Hence, only traceless matter (radiation) is supported by the field equations (\ref{feq}), (\ref{kg-eq}), which are derived from the action (\ref{tot-action}). This is due to the fact that the resulting theory is just Brans-Dicke theory with the anomalous value $\omega=-3/2$ of the BD coupling parameter. 

Let us to forget for a while about the above crucial deficiency of the theory (\ref{tot-action}), and assume anyway that matter degrees of freedom other than radiation and the conformally coupled scalars can be consistently included\footnote{The only possibility is to look for a convenient coupling of the matter Lagrangian to the dilaton.} in (\ref{feq}), (\ref{kg-eq}). The equation of motion of the matter degrees of freedom, properly the conservation of energy and stresses:

\bea \nabla^\kappa T^\text{mat}_{\kappa\mu}=0,\label{matt-cons}\eea is indeed transformed by (\ref{scale-t}). In other words, the equations of motion of matter (with non-vanishing trace) are not Weyl-invariant. Actually, under the conformal transformation of the metric \cite{faraoni-rev}: $g_{\mu\nu}\rightarrow\Omega^{-2}g_{\mu\nu}\;\Rightarrow$ 

\bea \nabla^\kappa T^\text{mat}_{\kappa\mu}=0\;\rightarrow\;\nabla^\kappa T^\text{mat}_{\kappa\mu}=-\frac{\der_\mu\Omega}{\Omega}\,T^\text{mat}.\label{cons-scale-t}\eea 

It is seen that the presence of matter with non-vanishing trace spoils any initially existing scale invariance. In this regard, the claim in \cite{bars''} -- to cite an example, but see also \cite{bars, bars', bars-ult} -- that the theory given by the following Lagrangian (Eq. (1) of Ref. \cite{bars''}):

\bea &&{\cal L}_\text{tot}=\frac{\left(\phi^2-\sigma^2\right)}{12}\,R+\frac{1}{2}(\der\phi)^2-\frac{1}{2}(\der\sigma)^2\nonumber\\
&&\;\;\;\;\;\;\;\;\;\;\;\;\;\;\;\;\;\;\;\;\;\;\;\;\;\;\;\;\;\;\;-\phi^4f(\sigma/\phi)+{\cal L}_\text{rad}+{\cal L}_\text{mat},\nonumber\eea is Weyl-invariant, is actually misleading. While the above claim were justified if remove the matter term ${\cal L}_\text{mat}$, in general it is incorrect. In \cite{bars} the argument used was that all particles masses $m$ are generated by the Higgs field -- in our notation it is the scalar field $\sigma$ corresponding to the non-vanishing component of the Higgs multiplet in the unitary gauge -- so that $m\propto\sigma(x)$. The authors conclude that due to this fact, since under (\ref{scale-t}) and $\sigma\rightarrow\Omega\,\sigma$; $m\rightarrow\Omega\,m$, $ds\rightarrow\Omega^{-1}ds$, the particle action $S=\int m\,ds$ is Weyl invariant. However, as we shall show in the next section, it is not enough that the action be explicitly scale invariant. Notice, additionally, that if remove the non-minimal coupling of the Higgs field to the curvature [$\sigma^2R\rightarrow 0$], the term $\propto-\phi^4f(\sigma/\phi)$, and perhaps other potential terms like the one $\propto(|\sigma|^2-v^2_0\phi^2)^2$, are invariant under (\ref{scale-t}) even if the resulting total action itself is not scale-invariant. In other words, the conformal coupling of $\sigma$ to the curvature $\sigma^2R+6(\der\sigma)^2$, which includes the non-minimal coupling $\propto \sigma^2R$, is cornerstone to allow for Weyl invariance. This means that even if the mass of particles in ${\cal L}_\text{mat}$ is proportional to the Higgs scalar $\sigma$ and the corresponding classical mechanics Lagrangian is Weyl invariant, a minimal coupling of matter to gravitation breaks the scale invariance of the theory.

In the next section, by means of geometric arguments, we shall show that even in the absence of matter the theory (\ref{deser-action}) -- the same for (\ref{bars-action}) -- is not actually a Weyl invariant setup.

\section{(pseudo)Riemann spacetimes}\label{riem}

Up to this point we have forgotten about the unavoidable discussion on the geometrical aspects of given gravitational theories being metric theories of gravity. In the present and subsequent sections we shall pay due attention to the geometrical aspects of the scale-invariant theories explored here. 

Although in the works of \cite{bars, bars', bars'', kallosh-linde, carrasco-kallosh, bars-ult, padilla, prester} no single word is said about the affine geometric structure of the spacetime manifold when discussing on conformal invariance, it is implicitly assumed that it is Riemann geometric theory which dictates the properties of the spacetime just like in general relativity. Likewise, in the present section we shall assume Riemannian spacetimes depicted by the pair $({\cal M},g_{\mu\nu})$, where ${\cal M}$ is the 4D manifold, and $g_{\mu\nu}$ the spacetime metric which obeys the Riemannian metricity condition

\bea \nabla_\mu g_{\nu\lambda}=0,\label{riem-metricity-c}\eea with the covariant derivative operator $\nabla_\mu$ defined in terms of $\{^\mu_{\nu\lambda}\}$ - the Christoffel symbols of the metric. 

Here we shall assume that the laws of gravity are governed by the vacuum action (\ref{deser-action}), i. e., no matter degrees of freedom other than the dilaton -- and perhaps other conformally coupled scalar fields like in (\ref{bars-action}) -- are considered. Under the conformal transformation of the metric in (\ref{scale-t}): $g_{\mu\nu}\rightarrow\Omega^{-2}g_{\mu\nu}$, the Christoffel symbols transform like \cite{faraoni-rev, faraoni}

\bea \{^\alpha_{\beta\sigma}\}\rightarrow\{^\alpha_{\beta\sigma}\}-\frac{1}{\Omega}\left(\delta^\alpha_\beta\der_\sigma\Omega+\delta^\alpha_\sigma\der_\beta\Omega-g_{\beta\sigma}\der^\alpha\Omega\right).\label{1-vp}\eea In this case it is understood that, given that Riemann geometry governs the affine properties of the original spacetime, the affine structure of the conformal space is also Riemannian, i. e., the Riemannian metricity condition $$\nabla_\mu g_{\nu\lambda}=0\;\rightarrow\;\nabla_\mu g_{\nu\lambda}=0,$$ is preserved by the conformal transformation of the metric, so that $({\cal M},g_{\mu\nu})\rightarrow({\cal M},g_{\mu\nu}).$ 

The point to notice is that, while the laws of gravity represented by (\ref{deser-action}) are unchanged by (\ref{scale-t}), the time-like Riemannian geodesics

\bea \frac{d^2x^\mu}{ds^2}+\{^\mu_{\kappa\nu}\}\frac{dx^\kappa}{ds}\frac{dx^\nu}{ds}=0,\label{riem-geod}\eea i. e., the motion equations for time-like point-particles in a curved spacetime $({\cal M},g_{\mu\nu})$  which solves (\ref{deser-action}), are mapped into non-geodesics of the conformal (also Riemannian) space $({\cal M},g_{\mu\nu})$:

\bea \frac{d^2x^\mu}{ds^2}+\{^\mu_{\kappa\nu}\}\frac{dx^\kappa}{ds}\frac{dx^\nu}{ds}=\frac{\der_\kappa\Omega}{\Omega}\frac{dx^\kappa}{ds}\frac{dx^\mu}{ds}-\frac{\der^\mu\Omega}{\Omega},\label{non-geod}\eea where under a convenient re-parametrization, $ds\rightarrow\Omega^{-1}d\tau$, the first term in the right-hand-side (RHS) above can be removed, but the second term $\propto-\Omega^{-1}\der^\mu\Omega$ can not be eliminated at all. Hence, given that (\ref{non-geod}) does not admit an affine parametrization whatsoever, this is not a Riemannian geodesic equation \cite{quiros-2013, wald}.

We are faced with the following situation: while in the original Riemannian spacetime $({\cal M},g_{\mu\nu})$ the time-like point-particles follow Riemannian geodesics (\ref{riem-geod}), in the conformal -- also Riemannian -- spacetime $({\cal M},g_{\mu\nu})$, the free-falling point-particles follow time-like curves which solve (\ref{non-geod}) which are not Riemannian geodesics. In both cases the laws of vacuum gravity are the same:

\bea &&G_{\mu\nu}=\frac{\lambda}{2}g_{\mu\nu}\phi^2-\frac{4}{\phi^2}\left[\der_\mu\phi\der_\nu\phi-\frac{1}{4}g_{\mu\nu}(\der\phi)^2\right]\nonumber\\
&&\;\;\;\;\;\;\;\;\;\;\;\;\;\;\;\;\;\;\;\;\;\;\;\;\;\;+\frac{2}{\phi}\left[\nabla_\mu\nabla_\nu\phi-g_{\mu\nu}\Box\phi\right],\label{vac-feq}\\
&&\Box\phi-\frac{1}{6}\,R\phi=\frac{\lambda}{3}\,\phi^3.\label{vac-kg-eq}\eea  This means that even in the vacuum -- supposedly scale invariant -- gravity (\ref{deser-action}), if probe the gravitational laws by means of test particles, the difference between gauges will be revealed.

The conclusion is trivial: since under the Weyl gauge transformations (\ref{scale-t}) the equations of motion of matter -- represented either by the conservation of stress-energy for matter fluxes or by the geodesic equations for point-like particles -- are indeed modified, the theory (\ref{deser-action}) (the same for (\ref{bars-action})) is not actually Weyl-invariant. Only the laws of gravity are unchanged by (\ref{scale-t}). As we shall see in the next section, this is a consequence of adopting spacetime manifolds whose geometrical structure is dictated by Riemann geometry, as the arena where to play the gravitational laws governed by (\ref{deser-action})/(\ref{bars-action}). The lesson to be learned is that there will be problems with a theory which pretends to be Weyl-invariant only because the action -- and the derived field equations -- is invariant under (\ref{scale-t}), but which is sustained by spacetimes whose geometrical structure does not share the gauge symmetry of the action.

\subsection{Gauge freedom}

The lack of Weyl gauge symmetry in the theory (\ref{deser-action}) is better illustrated by the following discussion. In the above vacuum field equations which are derived from (\ref{deser-action}), the KG equation (\ref{vac-kg-eq}) coincides with the trace of (\ref{vac-feq}). Hence, one degree of freedom is redundant:\footnote{We want to point out that even if consider radiation, in which case the Einstein equations (\ref{feq}) are satisfied, given that the radiation is traceless, the KG equation (\ref{kg-eq}) is also redundant since it coincides with the trace of the Einstein equations. This means that in this case a redundant degree of freedom to make Weyl gauge transformations (\ref{scale-t}) also arises.} there are -- in principle -- 10 equations to solve for 11 unknowns $g_{\mu\nu}$, $\phi$. The redundant degree of freedom is due to scale invariance of the action (\ref{deser-action}). What this means is that the vacuum field equations are not enough to determine the dynamics of the gravitational fields $(g_{\mu\nu},\phi)$. A feasible way out is to choose one specific gauge, say the GR gauge, or, as it is also called, the Einstein gauge (E-gauge) where $\phi_\text{E}=\sqrt{3/4\pi}\,G^{-1/2}$, with $G$- the Newton's constant. Then $g^\text{E}_{\mu\nu}$ is the metric which solves the Einstein-de Sitter equation $$G_{\mu\nu}=\frac{3\lambda}{8\pi G}\,g_{\mu\nu}.$$ The whole class of conformal spaces which solve (\ref{vac-feq}) can be then generated: 

\bea {\cal C}_\text{E}=\{(g^{(i)}_{\mu\nu},\phi_{(i)}):g^{(i)}_{\mu\nu}=\phi_{(i)}^{-2}g^\text{E}_{\mu\nu},\;\phi_{(i)}=f_i(x)\},\label{c-class}\eea where the $f_i$ are in the infinite countable set of the (non-vanishing) real-valued smooth functions. In other words, one has a whole (infinite) class of pairs $(g^{(i)}_{\mu\nu},\phi_{(i)})$ -- of spacetimes $({\cal M},g^{(i)}_{\mu\nu})$ correspondingly -- which satisfy the vacuum field equations (\ref{vac-feq}). The different spacetimes in the above class were equivalent geometrical representations of a given gravitational phenomenon if it were not for the 2nd term in the RHS of (\ref{non-geod}) which can be understood as an additional ``5-force'' $\propto-\Omega^{-1}\der^\mu\Omega$ of non-gravitational origin. Actually, given that in the E-gauge the test point-particles follow the Riemannian geodesics (\ref{riem-geod}), i. e., there is not additional 5-force in the E-gauge, in any of the conformal gauges in ${\cal C}_\text{E}$ given by Eq. (\ref{c-class}), the 5-force $\propto\phi_{(i)}^{-1}\der^\mu\phi_{(i)}$ arises which has different strength in the different conformal gauges. Hence, in principle, by doing experimentation with (time-like) test point-particles one is able to rule-out those gauges which do not meet the tight experimental bounds on 5-force \cite{will}. What this means is that (\ref{deser-action}) -- the same for (\ref{bars-action}) -- represents, in fact, a whole class of different theories related by Weyl gauge transformations (\ref{scale-t}). The same conclusion is obviously true for the theory (\ref{tot-action}) where matter degrees of freedom are considered.

\subsection{Spacetime singularities}\label{sing-sub}

The additional non-gravitational force in (\ref{non-geod}) is the responsible for the avoidance -- do not confound with removal -- of certain spacetime singularities which might be present in the original spacetime but which are not met in the conformal one: incomplete time-like geodesics in the original spacetime can be mapped into complete (non)geodesics in the conformal space thanks to the 5-force $\propto-\Omega^{-1}\der^\mu\Omega$, which deviates the motion of a time-like test particle from being geodesic. The role of the conformal transformations in this case is to send the singularity to one of the ends of the (complete) time-like non-geodesic curve at infinity. I. e., the singularity is still there but it takes an infinite proper (conformal) time along one such time-like non-geodesic curve to reach to it \cite{quiros-2000, kaloper}. Notice that we repeat ``time-like'' every time to emphasize that the given spacetime singularity may be hidden from time-like test observers but not from photons (in general from massless particles). Actually, the geodesic equations for massless particles are not affected by the conformal transformation of the metric \cite{wald}, so that the radiation -- this includes the gravitational waves -- always sees the singularity (see \cite{kaloper} for a related discussion). 

That the singularity which exists in one gauge is not removed in the conformal gauges can be seen if take a look at the Weyl gauge curvature invariants \cite{carrasco-kallosh}. One may argue that, working as we do with Riemannian spacetimes, it could be enough to explore the Riemannian invariants

\bea &&I_2\equiv R,\;I_4\equiv R_{\mu\nu}R^{\mu\nu},\;I'_4\equiv R_{\mu\nu\kappa\lambda}R^{\mu\nu\kappa\lambda},\nonumber\\
&&\;\;\;\;\;\;\;\;\;\;\;\;\;\;\;\;\;\;\;\;\;\;\;\;\;\;\;I''_4\equiv C^2:=C_{\mu\nu\kappa\lambda}C^{\mu\nu\kappa\lambda},\label{riem-inv}\eea where $C_{\mu\nu\kappa\lambda}$- the Weyl tensor, etc, but these are not Weyl gauge invariant quantities. Instead, the following curvature invariants

\bea I^\phi_2\equiv\phi^{-4}\left[\phi^2R+6(\der\phi)^2\right],\;I''^\phi_4\equiv\phi^{-4}C_{\mu\nu\kappa\lambda}C^{\mu\nu\kappa\lambda},\label{weyl-g-inv}\eea are actual Weyl gauge invariants. To construct curvature invariants related with the Ricci and the Riemann-Christoffel tensors is by far a more complex task \cite{kallosh-linde}.

By construction the gauge-free curvature scalars $I^\phi_2$ and $I''^\phi_4$ in (\ref{weyl-g-inv}) are invariant scalars both under general coordinate transformations and under Weyl gauge transformations (\ref{scale-t}), in other words, given that in the E-gauge, $$I^\phi_2=\phi^{-2}_\text{E}R(g_\text{E})=\frac{4\pi}{3}\,GR_\text{E},\;I''^\phi_4=\frac{16\pi^2}{9}\,G^2C^2_\text{E},$$ then 

\bea &&I^\phi_2=\frac{4\pi}{3}\,GR_\text{E}=\phi^{-4}\left[\phi^2R+6(\der\phi)^2\right],\nonumber\\
&&I''^\phi_4=\frac{16\pi^2}{9}\,G^2C^2_\text{E}=\phi^{-4}C^2.\nonumber\eea This means that if there is a curvature singularity at some spacetime point where $R_\text{E}$ and/or $C^2_\text{E}$ blow up in the E-gauge, this curvature singularity will be present in any other gauge. In Ref. \cite{carrasco-kallosh}, for instance, the Weyl gauge invariant $I''^\phi_4$ was used to show that any existing singularity in the E-gauge of the Weyl-invariant theory of \cite{bars, bars', bars''} will be also there in any of its Weyl conformal gauges (see a differing viewpoint in Ref. \cite{bars-ult}). As a matter of fact, given that the Weyl tensor governs the propagation of gravitational radiation through free space, this is a convenient quantity to judge about spacetime singularities in vacuum gravity. In a similar fashion, several years ago it was demonstrated in \cite{kaloper} that even if time-like free-falling observers in a conformal frame avoid a given existing singularity in the original frame, the null geodesics can not avoid hitting the singularity in a finite time since these are not modified by the conformal transformations. In other words, a given existing singularity can be hidden from time-like observers in the conformal frame, but it can not be hidden from massless particles (in \cite{kaloper} this was demonstrated for gravitational radiation).

Other gauge-free quantities may be constructed which are related with powers of scalar densities, instead of being just scalars. Take, for instance \cite{bars, bars', bars'', bars-ult, kallosh-linde, carrasco-kallosh}: $${\cal J}_0\equiv |g|^{1/4}\frac{\phi^2-\sigma^2}{6}.$$ In this case a power of the square of the metric determinant $\sqrt{|g|}$ is implied. This is a scalar density of weight $+1$: $\sqrt{|g'|}=J\sqrt{|g|}$, where $J=\det[\der x^\lambda/\der x'^\kappa],$ is the Jacobian determinant. It is included as part of the measure in the action integral in combination with the volume element: $d^4x\sqrt{|g|}$, to allow for general coordinate invariance. Since under general coordinate transformations the scalar densities are in general transformed, the utility of gauge-invariant quantities like ${\cal J}_0$ in the study of the spacetime singularities is unclear. To illustrate this let us assume the FRW metric with flat spatial sections. Given the peculiar properties of the cosmological metric, it is $dt\sqrt{|g|}$ which transforms like $\sqrt{|g|}$ for a general metric, under a conformal transformation $g_{\mu\nu}\rightarrow\Omega^{-2}g_{\mu\nu}$. Actually, let us write the FRW (flat) metric in the Einstein frame and in any other conformal frame $g^E_{\mu\nu}=\Omega^2g_{\mu\nu}$: $$ds_\text{E}^2=-dt_\text{E}^2+a_\text{E}^2(t_\text{E})d{\bf x}^2,\;ds^2=-dt^2+a^2(t)d{\bf x}^2,$$ where $$dt=\Omega^{-1}dt_\text{E},\;a(t)=\Omega^{-1}a_\text{E}(t_\text{E}),$$ then $$\sqrt{|g|}=\Omega^{-3}\sqrt{|g_E|},\;dt\sqrt{|g|}=\Omega^{-4}dt_E\sqrt{|g_E|}.$$ 

Misunderstanding of this fact may lead to the following situation. Let us assume a single conformally coupled scalar field like in (\ref{tot-action}). For the flat FRW metric we have that ${\cal J}_0=a^{3/2}\phi^2/6.$ It is known that the action (\ref{tot-action}) is mapped into the GR Einstein-Hilbert action if choose $\Omega=\phi/\phi_E$, where, for simplicity, we assume that the constant $\phi_E=1$. Hence $dt=\phi^{-1}dt_\text{E}$, $a(t)=\phi^{-1}a_\text{E}(t_\text{E})$. Since $\sqrt{|g_E|}=a_E^3$, from the definition above we have that (recall that $\phi_E=1$) $${\cal J}^E_0=a_E^{3/2}/6,\;{\cal J}_0=a^{3/2}\phi^2/6.$$ If assume that ${\cal J}_0$ is an actual invariant as it is done in \cite{bars, bars', bars'', bars-ult}, then $${\cal J}^E_0={\cal J}_0\;\Rightarrow\;a=\phi^{-4/3}a_E,$$ which is incorrect. It is obvious that ${\cal J}_0$ is a Weyl-gauge invariant, but, as said, it is not an invariant under general coordinate transformations, and in the above equations, due to the peculiar properties of the cosmological metric, we have taken into account that $dt=\phi^{-1}dt_E$, which is a coordinate transformation. Besides, if assume the theory (\ref{tot-action}) with $\lambda=0$ and ${\cal L}_\text{mat}={\cal L}_\text{rad}$ -- the Lagrangian for radiation to govern the cosmological dynamics of flat FRW spacetimes, then, since $3H^2_E=\rho_\text{rad}\propto a^{-4}_E$ $\Rightarrow\;a_E\propto t_E^{1/2}$, there is a real cosmological singularity at $t_E=0$, where $\rho_\text{rad}\propto t^{-2}_E$ blows up, even if $${\cal J}_0=a^{3/2}_E\phi^2_E=a^{3/2}_E\propto t^{3/4}_E,$$ is finite at $t_E=0$.

Likewise the quantity ${\cal J}_0$ one may invent many other such gauge-invariant combinations like, for instance,

\bea &&{\cal J}_1\equiv\sqrt{|g|}\left[\frac{\left(\phi^2-\sigma^2\right)}{12}\,R+\frac{1}{2}(\der\phi)^2-\frac{1}{2}(\der\sigma)^2\right],\nonumber\\
&&{\cal J}_2\equiv\sqrt{|g|}C_{\mu\nu\kappa\lambda}C^{\mu\nu\kappa\lambda},\nonumber\eea etc, but, what is the point of this? We have genuine gauge-free curvature invariants (scalars) like $I^\phi_2$ and $I''^\phi_4$ above, which may decide whether a given singularity is a true curvature singularity or just a coordinate or gauge artifact (see the discussion in section \ref{sing}).

In the next sections we shall explore a feasible slight modification of the Riemann spacetime structure in the search for a different geometrical arena where to play the scale-invariant laws of gravity. It will be shown, in particular, that the search for gauge-independent curvature invariants related with the curvature scalar, and with the Ricci, Riemann and Weyl tensors is by far a much more simple task.

\section{Weyl-integrable geometry}\label{wing}

In the former sections we have assumed -- as it is done usually in the bibliography on scale invariance (see, for instance \cite{bars, bars', bars'', kallosh-linde, carrasco-kallosh, bars-ult, padilla, percacci, prester, shaposhnikov}) -- that the affine properties of the underlying spacetimes are governed by Riemann geometry. However, in Riemannian spacetimes, given the transformation properties of the Riemannian affine connection (properly the Christoffel symbols of the metric) depicted by (\ref{1-vp}), the affine properties are not preserved by the Weyl gauge transformations (\ref{scale-t}).

Let us look for feasible modifications of Riemann geometry which could accommodate Weyl gauge symmetry. The first (and simplest) such modification that comes to one's mind is Weyl geometry \cite{weyl, perlick, novello, scholz-h, bib-weyl, books, chinos, chinos-1}. Weyl's geometric theory is a minimal generalization of Riemann geometry to include point dependent length of vectors during parallel transport, in addition to the point dependent property of vectors directions.\footnote{In Ref. \cite{chinos} the conformal invariance in Einstein-Cartan-Weyl spaces is investigated, while in \cite{chinos-1} Weyl invariance is extended to anisotropic spacetime version.} It is assumed that the length of a given vector ${\bf l}$ ($l\equiv\sqrt{g_{\mu\nu} l^\mu l^\nu}$) which is submitted to parallel transport varies from point to point in spacetime according to: $dl=l w_\mu dx^\mu/2$, where $w_\mu$ is the Weyl gauge boson. Here we shall concentrate in a special branch of Weyl geometry dubbed as Weyl-integrable geometry or WING for short. WING is obtained from Weyl geometry by replacing $w_\mu\rightarrow\der_\mu w$, where $w$ is the Weyl gauge scalar.\footnote{In this case, since $\oint dx^\mu \der_\mu w/2=0$, then the lengths of vectors, although point-dependent, are integrable. The units of measure in WING manifolds are point-dependent quantities unlike in Riemannian spaces. For a concise exposition of the fundamentals of Weyl-integrable geometry we submit the reader to the classical texts \cite{books}.} 

Here we shall consider the action (\ref{deser-action}), and we will postulate that the spacetimes which solve the derived field equations have WING affine structure in place of the Riemannian structure assumed in the former sections. In order to make the theory (\ref{deser-action}) compatible with the former postulate we shall assume that the dilaton $\phi$ is the Weyl gauge scalar -- the precise correspondence with the well-known Weyl boson is actually $w=\ln\phi^2$ -- so that it is lifted to the category of a geometric field in addition to the metric field itself $g_{\mu\nu}$. The corresponding WING affine connection of the manifold $\Gamma^\alpha_{\beta\mu}$ is defined as 

\bea \Gamma^\alpha_{\beta\mu}\equiv\{^\alpha_{\beta\mu}\}+\frac{1}{\phi}\left(\delta^\alpha_\beta\der_\mu\phi+\delta^\alpha_\mu\der_\beta\phi-g_{\beta\mu}\der^\alpha\phi\right).\label{wig-aff-c}\eea The resulting Weyl-invariant action -- which is coupled to WING background spacetimes -- reads 

\bea S^{(w)}=\frac{1}{12}\int d^4x\sqrt{|g|}\left[\phi^2R^{(w)}+\lambda\phi^4\right],\label{deser-mod}\eea where the label ``$(w)$'' refers to Weyl-integrable quantities, which are defined with respect to the affine connection (\ref{wig-aff-c}). Notice that the kinetic term for the Weyl gauge boson (former dilaton) $\phi$ is already included in the definition of the WING-curvature scalar (up to a divergence) $$R^{(w)}=R+6\frac{(\der\phi)^2}{\phi^2},$$ where in the RHS of this equation the given quantities and operators coincide with their Riemannian definitions in terms of the Christoffel symbols of the metric.

One may as well consider adding a non-geometric, minimally coupled scalar field $\sigma$, which under (\ref{scale-t}) transforms like $\sigma\rightarrow\Omega\sigma$. The resulting Weyl invariant action can be written as follows (here we omit writing the quartic potential term $\propto\phi^4$):

\bea S^{(w)}=\int d^4x\sqrt{|g|}\left\{\frac{1}{12}\,\phi^2R^{(w)}-\frac{1}{2}(D\sigma)^2\right\},\label{deser-mod'}\eea where, as before, the different geometric quantities and operators labeled with the ``$(w)$'' refer to WING objects, and the standard derivative of the non-geometric field minimally coupled to gravity $\sigma$, have been replaced by the gauge-covariant derivative $$\der_\mu\sigma\rightarrow D_\mu\sigma\equiv\left(\der_\mu-\frac{\der_\mu\phi}{\phi}\right)\sigma.$$ Following this procedure one might add any number of minimally coupled scalars. 

This apparently slight modification of (\ref{deser-action}) designed to make that theory compatible with WING backgrounds, results in that scale invariance is an actual symmetry of the laws of gravity. First of all notice that the torsionless affine connection of the WING space (\ref{wig-aff-c}), as well as the non-metricity condition 

\bea \nabla^{(w)}_\mu g_{\kappa\lambda}=-2\frac{\der_\mu\phi}{\phi}\,g_{\kappa\lambda}\;\Rightarrow\;\nabla_{(w)}^\kappa g_{\kappa\lambda}=-2\frac{\der^\kappa\phi}{\phi}\,g_{\kappa\lambda},\label{wing-law}\eea which is the fundamental geometric law of WING spaces, both are unchanged by the scale transformations (\ref{scale-t}). As a consequence, the WING-Ricci tensor $R^{(w)}_{\mu\nu}$ and the corresponding WING-Einstein's tensor $G^{(w)}_{\mu\nu}\equiv R^{(w)}_{\mu\nu}-g_{\mu\nu} R^{(w)}/2$, as well as the covariant derivative operator $\nabla^{(w)}_\mu$, etc, are scale-invariant objects.

In this Weyl-invariant modification of (\ref{deser-action}) which is grounded in WING backgrounds, not only the field equations derived from (\ref{deser-mod})/(\ref{deser-mod'}) -- see below -- but also the WING-geodesics $$\frac{d^2x^\alpha}{ds^2}+\Gamma^\alpha_{\mu\nu}\frac{dx^\mu}{ds}\frac{dx^\nu}{ds}=\frac{\der_\mu\phi}{\phi}\frac{dx^\mu}{ds}\frac{dx^\alpha}{ds},$$ or after a convenient re-parametrization $d\sigma=\phi\,ds$, 

\bea \frac{d^2x^\alpha}{d\sigma^2}+\Gamma^\alpha_{\mu\nu}\frac{dx^\mu}{d\sigma}\frac{dx^\nu}{d\sigma}=0,\label{mod-geod}\eea and the WING continuity equation 

\bea \nabla_{(w)}^\kappa T^\text{mat}_{\kappa\mu}=2\frac{\der^\kappa\phi}{\phi}\,T^\text{mat}_{\kappa\mu},\label{mod-cons-eq}\eea all are invariant under the Weyl gauge transformations (\ref{scale-t}). In equation (\ref{mod-cons-eq}) -- compare with the non-metricity condition in the form of the right-hand side Eq. (\ref{wing-law}) -- the term in the RHS is not actually a source term, instead it expresses the fact that in WING spaces the units of measure of energy and stresses are point-dependent quantities, as well as the length of any other vector. As a matter of fact the above conservation equation looks similar to that in GR if redefine the stress-energy tensor of matter $T^{w,\text{mat}}_{\mu\nu}\equiv\phi^{-2}T^\text{mat}_{\mu\nu}$,

\bea \nabla_{(w)}^\kappa T^{w,\text{mat}}_{\kappa\mu}=0.\label{wig-cont-eq}\eea This is the stress-energy tensor which has the Weyl-invariant physical meaning since, under (\ref{scale-t}), it is unchanged like the WING-Einstein's tensor $G^{(w)}_{\mu\nu}$.

\subsection{Field equations}

Let us to explore an specific example. Consider the following action \cite{quiros-2014}:

\bea &&S^{(w)}_\text{example}=\int d^4x\sqrt{|g|}\left\{\frac{1}{12}\,\phi^2R^{(w)}-\right.\nonumber\\
&&\left.\;\;\;\;\;\;\;\;\;\;\;\;\;\;\;\;\;\;\;\;\;\;\;\frac{1}{2}|D\sigma|^2-\frac{\lambda}{4}\left(|\sigma|^2-v^2_0\phi^2\right)^2\right\},\label{ex-action}\eea where $\sigma$ can be the Higgs field in the unitary gauge, in which case $v_0$ is the mass parameter of the standard model of particles, and $\lambda$ is a coupling constant. The above action is invariant under (\ref{scale-t}), plus the Higgs field transformation $\sigma\rightarrow\Omega\,\sigma$. In order to derive the field equations it is convenient to introduce a gauge-independent scalar field variable $\chi=\phi^{-1}\sigma$. After this the action (\ref{ex-action}) is rewritten as

\bea S^{(w)}_\text{example}=\int d^4x\sqrt{|g|}\left\{\frac{1}{12}\,\phi^2R^{(w)}+\phi^2{\cal L}_\text{Higgs}\right\},\label{chi-action}\eea where the Higgs field Lagrangian is given by

\bea {\cal L}_\text{Higgs}=-\frac{1}{2}(\der\chi)^2-\frac{\lambda\phi^2}{4}(\chi^2-v^2_0)^2.\label{higgs-lag}\eea 

Written in the form (\ref{chi-action}) the action of the theory is manifestly Weyl-invariant, where we recall that under the scale transformations (\ref{scale-t}) the field $\chi$ is unchanged, i. e., it is already a Weyl gauge invariant scalar. Notice also that, unlike in (\ref{deser-action})/(\ref{bars-action}) the minimal coupling of the Higgs field to gravity does not spoil the scale invariance of the action. The following manifestly scale-invariant field equations are easily derived from (\ref{chi-action}):

\bea &&G_{\mu\nu}^{(w)}=\frac{1}{M^2_\text{pl}}\,T^{(w,\chi)}_{\mu\nu}\;\Rightarrow\nonumber\\
&&G_{\mu\nu}-\frac{2}{\phi}(\nabla_\mu\der_\nu\phi-g_{\mu\nu}\Box\phi)+\nonumber\\
&&\;\;\;\;\;\;\;4\frac{\der_\mu\phi}{\phi}\frac{\der_\nu\phi}{\phi}-g_{\mu\nu}\frac{(\der\phi)^2}{\phi^2}=\frac{1}{M^2_\text{pl}}\,T^{(w,\chi)}_{\mu\nu},\label{e-feq}\\
&&\nabla^\kappa_{(w)}T^{(w,\chi)}_{\kappa\mu}=0\;\Rightarrow\nonumber\\
&&\Box\chi+2\frac{(\der\phi\cdot\der\chi)}{\phi}=\lambda\phi^2\left(\chi^2-v^2_0\right)\chi,\label{chi-kg-eq}\eea where in the second lines (explicit form) of the above equations the geometric objects and operators are the usual Riemannian quantities, $M^2_\text{pl}=1/6$ in the units chosen in this paper, and

\bea &&T^{(w,\chi)}_{\mu\nu}=-\frac{2}{\sqrt{|g|}}\frac{\der(\sqrt{|g|}\,{\cal L}_\text{Higgs})}{\der g^{\mu\nu}}\nonumber\\
&&\;\;\;\;\;\;\;\;\;\;\;=\der_\mu\chi\der_\nu\chi-\frac{1}{2}g_{\mu\nu}(\der\chi)^2\nonumber\\
&&\;\;\;\;\;\;\;\;\;\;\;\;\;\;\;\;\;\;\;\;\;\;\;\;\;\;\;\;-\frac{\lambda\phi^2}{4}\,g_{\mu\nu}\left(\chi^2-v^2_0\right)^2,\label{chi-set}\eea is the already Weyl-invariant stress-energy tensor of the Higgs field. If derive the KG equation for the Weyl gauge scalar $\phi$ (former dilaton) from the action (\ref{chi-action}), one obtains $$\Box\phi-\frac{1}{6}R\phi=\phi\,T^{(w,\chi)}=-\phi\left[(\der\chi)^2+\lambda\phi^2\left(\chi^2-v^2_0\right)^2\right],$$ which exactly coincides with the trace of the Einstein's equations (\ref{e-feq}) independent of the type of matter under consideration. Recall that in the standard situation in the references \cite{bars, bars', bars'', kallosh-linde, carrasco-kallosh, bars-ult, padilla, percacci, prester, shaposhnikov} where Riemannian backgrounds are adopted, only traceless matter can be consistently added to the action (\ref{deser-action}) (same for (\ref{bars-action})).

Summarizing: the adoption of Weyl-integrable geometry as the theory which governs the affine properties of the spacetime background allows to construct a fully Weyl-invariant theory even if matter degrees of freedom other than radiation (and non-minimally coupled scalars) are considered. The subtle idea was to lift the dilaton $\phi$ to the category of a geometric field, i. e., to adopt it as the Weyl gauge boson $w=\ln\phi^2$ of WING, i. e., the one which takes part in the definition of the affine connection (\ref{wig-aff-c}) and consequently in the definition of the curvature of spacetime.

\subsection{Gauge freedom}

As already mentioned, the KG equation for the Weyl gauge boson $\phi$ is not an independent equation, but it coincides with the trace of the WING-Einstein equations (\ref{e-feq}). This is a direct consequence of scale invariance which means that there is not just a single Weyl-integrable space $({\cal M},g_{\mu\nu},\phi)$ -- properly a gauge -- which is solution of (\ref{e-feq}), (\ref{chi-kg-eq}), but a whole equivalence class of them: 

\bea &&{\cal C}=\{(g^{(a)}_{\mu\nu},\phi_{(a)}):a=1,2,...,i,...,k,...,\infty;\nonumber\\
&&\;\;\;\;\;\;\;\;\;\;\;\;\;\;\;\;\;\;\;\;\;g^{(i)}_{\mu\nu}=\Omega^2_{ik}g^{(k)}_{\mu\nu},\;\phi_{(i)}=\Omega^{-1}_{ik}\phi_{(k)}\},\nonumber\eea i. e., any two pairs in ${\cal C}$: $(\bar g_{\mu\nu},\bar\phi)$ and $(g_{\mu\nu},\phi)$, are related by a scale transformation $\bar g_{\mu\nu}=\Omega^2g_{\mu\nu}$, $\bar\phi=\Omega^{-1}\phi$. As it was illustrated in the cosmological context in \cite{quiros-2014}, the most one can get from the field equations is a functional relationship among the gravitational potentials $g_{\mu\nu}$ and $\phi$. This relationship is independent of the matter content: for a given matter source one have an infinity of possibilities $(g^{(a)}_{\mu\nu},\phi_{(a)})$, where $a=1,2...,\infty$ and the $\phi_{(a)}$ belong in the space of continuous real-valued functions. Each possible gauge $(g^{(i)}_{\mu\nu},\phi_{(i)})\in{\cal C}$, represents a potential geometric description of the laws of gravity (and of particle's physics). From the cosmological standpoint, for instance, to have an infinity of feasible -- fully equivalent -- geometrical descriptions amounts to have an infinity of possible patterns of cosmological evolution which satisfy the same cosmological field equations. For a detailed discussion of this issue and possible connections with the multiverse picture \cite{multiverse} see \cite{quiros-2014}.

The simplest gauge one may choose is the one where $\phi=\phi_\text{E}=$ const. This trivial E-gauge corresponds to general relativity since, after the above choice, the action (\ref{ex-action}) transforms into the Einstein-Hilbert (EH) action minimally coupled to the standard model of particles with no new physics beyond the standard model at low energies. At the same time WING spaces are transformed into pseudo-Riemannian manifolds. 

It is a very simple exercise to show that the spacetimes which solve the GR field equations belong in the conformal class ${\cal C}$. Actually, let us consider the WING-EH action (\ref{deser-mod}) $$S^{(w)}_\text{EH}=\frac{1}{12}\int d^4x\sqrt{|g|}\left[\phi^2R^{(w)}+\lambda\phi^4\right].$$ Under a Weyl rescaling (\ref{scale-t}) with $\Omega^2=\phi^2$, the WING affine connection (\ref{wig-aff-c}) is transformed into the Christoffel symbols of the conformal metric: $$\Gamma^\mu_{\alpha\beta}\rightarrow\{^\mu_{\alpha\beta}\}\;\Rightarrow\;R^{(w)}_{\mu\nu}\rightarrow R_{\mu\nu},\;\text{etc.}$$ then $$S^{(w)}_\text{EH}\rightarrow S_\text{EH}=\frac{1}{12}\int d^4x\sqrt{|g|}\left[R+\lambda\right].$$ Through an inverse transformation (\ref{scale-t}) with $\Omega^2=\phi^{-2}$, each GR solution generates an infinite set of spacetimes back in ${\cal C}$. To see this let us start with a given known GR solution $({\cal M},g^\text{E}_{\mu\nu})$, and perform the conformal transformation of the GR metric: $$g^\text{E}_{\mu\nu}\rightarrow\Omega^2g^\text{E}_{\mu\nu},\;\Omega^2=\phi^{-2}.$$ Under the above transformation, the Riemannian quantities and operators are mapped back into WING objects: $$\{^\mu_{\alpha\beta}\}\rightarrow\Gamma^\mu_{\alpha\beta}\;\Rightarrow\;R_{\mu\nu}\rightarrow R^{(w)}_{\mu\nu},\;\text{etc.}$$ 

It is not difficult to realize that there exists an infinite countable set of smooth real-valued functions $\phi_{(a)}=f_a(x)$, such that every pair $$(g^{(a)}_{\mu\nu},\phi_{(a)}),\;\text{with}\;g^{(a)}_{\mu\nu}=\phi_{(a)}^{-2}g^\text{E}_{\mu\nu},\;a=1,2,...\infty,$$ belongs in ${\cal C}$. This means that each GR spacetime solution $({\cal M},g^\text{E}_{\mu\nu})$, together with its infinite set of equivalent conformal representations $$\{({\cal M},g^{(a)}_{\mu\nu},\phi_{(a)}):\,g^{(a)}_{\mu\nu}=\phi_{(a)}^{-2}g^\text{E}_{\mu\nu},\;a=1,2,...\infty\},$$ belong in the equivalence class ${\cal C}$.

\section{Gauge-free curvature invariants and spacetime singularities}\label{sing}

There are issues which require of gauge-independent geometric invariants to reach to gauge-independent conclusions. An outstanding example is the singularity issue. There has been a debate about the possibility that certain spacetime singularities in scalar-tensor theories can be avoided in their conformal formulations \cite{quiros-2000, faraoni, kaloper}. Unlike scalar-tensor theories which are not scale invariant, the theory (\ref{ex-action}) is Weyl gauge-invariant so that the discussion of this issue is clearly different. 

In order to discuss on the occurrence of spacetime singularities in the above setup one is obliged to resort to the geometric invariants like

\bea &&I^\phi_2\equiv\phi^{-2}R^{(w)},\;I^\phi_4=\phi^{-4}R^{(w)}_{\mu\nu}R^{\mu\nu}_{(w)},\nonumber\\
&&I'^\phi_4\equiv\phi^{-4}R^{(w)}_{\mu\nu\kappa\lambda}R^{\mu\nu\kappa\lambda}_{(w)},\label{wing-inv}\eea etc, which are not transformed by the scale transformations (\ref{scale-t}) and, hence, are the ones which carry gauge-independent physical meaning. 

As it was demonstrated in the former section, general relativity is a particular gauge of the theory (\ref{ex-action}) when the Weyl scalar $\phi=\phi_\text{E}$ is a constant where, without lost of generality, we may set $\phi_\text{E}=1$. This entails that the following equalities involving the gauge-independent curvature invariants (\ref{wing-inv}) are satisfied:

\bea &&I^\phi_2=R_\text{E}=\phi^{-2}R^{(w)},\nonumber\\
&&I^\phi_4=R^\text{E}_{\mu\nu} R_\text{E}^{\mu\nu}=\phi^{-4}R^{(w)}_{\mu\nu}R^{\mu\nu}_{(w)},\nonumber\\
&&I'^\phi_4=R^\text{E}_{\mu\nu\kappa\lambda} R_\text{E}^{\mu\nu\kappa\lambda}=\phi^{-4}R^{(w)}_{\mu\nu\kappa\lambda}R^{\mu\nu\kappa\lambda}_{(w)},\label{sing-inv}\eea where the quantities with an ``E'' denote Riemannian objects defined in terms of the Christoffel symbols of the metric $g^\text{E}_{\mu\nu}$ in the E-gauge. Notice that equalities like the ones in (\ref{sing-inv}) arise only in gauge invariant theories where the gauge-independent curvature invariants (\ref{wing-inv}) make sense. These are not necessarily satisfied when dealing with standard scalar-tensor theories like the BD theory \cite{bd}. 

Let us assume the following hypothetical situation: the spacetime in the E-gauge $({\cal M},g^\text{E}_{\mu\nu})$ has a curvature singularity at some point $x_P$ in the manifold, such that the GR invariant $$I'^\phi_4=R^\text{E}_{\mu\nu\kappa\lambda} R_\text{E}^{\mu\nu\kappa\lambda}=\alpha(x),$$ blows up at $x_P$, i. e., 

\bea \lim_{x\rightarrow x_P}\alpha(x)\rightarrow\infty.\label{lim}\eea Let us further assume that $$\lim_{x\rightarrow x_P}\phi^4\rightarrow 0,$$ where the latter limit is approached in such a way that the WING-curvature invariant $$R^{(w)}_{\mu\nu\kappa\lambda}R^{\mu\nu\kappa\lambda}_{(w)}=\alpha\phi^4,$$ remains finite at $x_P$. This means that, given that the conditions of the hypothetical situation described above are fulfilled, the curvature singularity in the E-gauge is not felt by an observer living in the conformal WING-world. This does not mean that the singularity has been erased by the Weyl rescaling

\bea g_{\mu\nu}=\Omega^{-2}g^\text{E}_{\mu\nu},\;\phi=\Omega\phi_\text{E}.\label{scale-t'}\eea As a matter of fact the physically meaningful gauge-independent curvature invariant $I'^\phi_4=\alpha$ blows up at $x_P$, meaning that the singularity is still there. To understand what actually happens when the singularity is approached one have to recall that in WING spacetimes the units of measure are point-dependent. As the singularity is approached and the given gauge-independent curvature invariant, say $I'^\phi_4$, grows up without bound, the corresponding units of measure increase in a similar fashion so that the increase in the magnitude of the invariants is conveniently balanced.

\subsection{Big-gang singularity}

For a more specific qualitative analysis it is convenient to study a concrete example. Here we shall consider as an specific example the GR cosmological singularity usually associated with the big-bang. We have $$ds_\text{E}^2=-dt_\text{E}^2+a_\text{E}^2(t_\text{E})d{\bf x}^2,\;ds^2=-dt^2+a^2(t)d{\bf x}^2,$$ where the left-hand line-element refers to the (conformal) GR Friedmann-Robertson-Walker (FRW) spacetime, while the right-hand one refers to WIG-FRW spacetime. The following relationships arise: 

\bea dt=\phi^{-1}dt_\text{E},\;a(t)=\phi^{-1}a_\text{E}(t_\text{E}),\;\rho^{(w)}=\phi^2\rho_\text{E},\label{rel}\eea where $\rho^{(w)}$ is the energy density of matter measured by a co-moving observer in the WING-world, and we assumed the conformal factor $\Omega^2=\phi^2$. Consider a singular GR solution $a_\text{E}\propto t_\text{E}^n$ ($n$ is an arbitrary positive constant), so that according to the GR-Friedmann constraint [$3H_\text{E}^2=\rho_\text{E}/M^2_\text{pl}$], we have $\rho_\text{E}\propto t_\text{E}^{-2}$. The initial cosmological singularity is at $t_\text{E}=0$ where the matter energy density $\rho_\text{E}$ blows up. It is a simple exercise to show that invariants $I^\phi_2$, $I^\phi_4$ and $I'^\phi_4$ go to infinity at the singularity. In particular $I^\phi_2\propto H_\text{E}^2\propto t_\text{E}^{-2}$ grows up without bound as $t_\text{E}\rightarrow 0$. Among the infinity of possibilities let us choose\footnote{Recall that due to the gauge freedom we are free to choose any $\phi$ we want.} 

\bea \phi(t_\text{E})=\tanh(t_\text{E}).\label{vphi}\eea After the above convenient choice one gets that the WING curvature scalar $$R^{(w)}=\phi^2 I^\phi_2\propto\tanh^2(t_\text{E})/t^2_\text{E},$$ is always bounded: $0\leq t_\text{E}<\infty$ $\Rightarrow\,1\geq R^{(w)}>0$. The same is true for the energy density measured by a co-moving WING-observer $\rho^{(w)}=\phi^2\rho_\text{E}\propto R^{(w)}$. Hence a co-moving observer in the equivalent WING picture does not find singular behavior at all. As already stated, the explanation is simple: although the singularity is still there, as the co-moving observer approaches to it, any unbounded increment in any of the gauge-independent curvature invariants is balanced by a proportional increment in the corresponding units of measure in the WING-FRW spacetime.

Regarding the amount of cosmic time separating a given co-moving observer from the singularity we have to say that, while the interval of cosmic time from, say, the present moment of the cosmic history $t^\text{E}_0=t_0$ to the singularity in the past at $t_\text{E}=0$, is finite, in terms of the cosmic time measured by an observer in the WING-world we have that $$\Delta t=\int_\epsilon^{t_0}\frac{dt_\text{E}\cosh(t_\text{E})}{\sinh(t_\text{E})}=\ln\left(\frac{\sinh(t_0)}{\sinh(\epsilon)}\right),$$ where $\epsilon$ is a small number such that, as one approaches to the initial singularity, $\epsilon\rightarrow 0$. Hence, as $\epsilon\rightarrow 0$, $\Delta t\rightarrow\infty$. This means that to an observer in the WING-world the initial (big-bang) singularity is an infinite amount of cosmic time into the past. In consequence, for all practical purposes the corresponding WIG geodesics are complete into the past and the initial singularity -- although not erased -- is effectively avoided. Recall, however, that null-geodesics always meet the singularity in a finite time.

\section{Discussion and Conclusion}\label{conclu}

Although most theories of the fundamental interactions (including general relativity and string theory) assume that the geometric structure of spacetime is pseudo-Riemann, there are indications that a generalization of Riemann geometry -- notably Weyl integrable geometry -- might represent a better suited arena where to formulate the laws of gravity \cite{quiros-2013, quiros-2014, dicke, smolin, cheng, scholz}. As a matter of fact, it is surprising that Weyl geometry -- the natural arena for Weyl invariance -- is not usually adopted when dealing with scale invariance of the laws of gravity. Instead -- and regrettably -- any kind of discussion on the geometrical structure of spacetime is avoided when dealing with Weyl invariance. See, for instance, the incomplete list of references \cite{bars, bars', bars'', kallosh-linde, carrasco-kallosh, bars-ult, padilla, percacci, prester, shaposhnikov, deser} for a clear illustration of this statement. 

Here we have learned that it is not enough to postulate a Weyl-invariant action like, for instance (see equations (\ref{deser-action}) and (\ref{bars-action})): $$S=\int d^4x\sqrt{|g|}\left[\frac{\phi^2}{12}\,R+\frac{1}{2}(\der\phi)^2\right],$$ or $$S=\int d^4x\sqrt{|g|}\left[\frac{\left(\phi^2-\sigma^2\right)}{12}\,R+\frac{1}{2}(\der\phi)^2-\frac{1}{2}(\der\sigma)^2\right],$$ which are invariants under the Weyl gauge transfromations $$g_{\mu\nu}\rightarrow\Omega^{-2}g_{\mu\nu},\;\phi\rightarrow\Omega\phi,\;\sigma\rightarrow\Omega\sigma.$$ Additionally one has to make a separate postulate on the geometrical laws which govern the affine structure of the background spacetime. If one postulates Riemann geometry -- as it is usually done --, since the Riemannian affine structure is modified by the Weyl gauge transformations, then scale invariance is condemned to fail. If, alternativelly, postulate that the affine properties of the space are dictated by Weyl-integrable geometry, then there is room for scale invariance to be an actual symmetry of the laws of gravity \cite{quiros-2013, quiros-2014}.

In section \ref{anomal} we have shown that, if associated with Riemannian spacetime backgrounds, the above actions (\ref{deser-action}) and (\ref{bars-action}) just coincide with Brans-Dicke theory with the anomalous value of the BD coupling constant $\omega=-3/2$. As it is well-known, in such a case only conformally coupled scalars and radiation do not explicitly destroy the Weyl invariance. We wrote ``explicitly'' because, as a matter of fact, even the vacuum theories depicted by these kinds of action are actually not conformally invariant theories. This is demonstrated by the fact that the motion equations for time-like test particles are not invariant under the Weyl gauge transformations. Of course, this is a consequence of the wrong choice of the laws of geometry: Riemannian geodesics are not invariant under the conformal transformations of the metric. 

That misunderstanding of this fact may lead to wrong conclusions can be illustrated with the discussion on geodesic completeness within theories given by the action (\ref{bars-action}) (second equation at the beginning of this section) \cite{bars, bars', bars'', carrasco-kallosh, bars-ult}. In \cite{bars, bars', bars'', bars-ult} the authors rely on the study of time-like geodesics for particles with the mass $m$. They then made the argument that going to the limit of vanishing mass the completeness of null-geodesics can be incorporated in the discussion. Apparently, these authors forgot that their action is just BD gravity with BD coupling parameter $\omega=-3/2$, in which case only matter in the form radiation can be consistently considered, where by ``consistently'' we mean ``consistent with the derived field equations.'' A simple check of the field equations in their theory will clearly reveal that only radiation can be considered (see the related discussion in section \ref{anomal} of this paper). What is more disturbing: even if consider the vaccum theory, given that the background spacetimes are of Riemann affine structure -- which is actually modified by the conformal transformations -- the equations of motion of time-like test particles are transformed by the Weyl gauge transformations:

\bea &&\frac{d^2x^\mu}{ds^2}+\{^\mu_{\kappa\nu}\}\frac{dx^\kappa}{ds}\frac{dx^\nu}{ds}=0\;\rightarrow\nonumber\\
&&\frac{d^2x^\mu}{ds^2}+\{^\mu_{\kappa\nu}\}\frac{dx^\kappa}{ds}\frac{dx^\nu}{ds}=\frac{\der_\kappa\Omega}{\Omega}\frac{dx^\kappa}{ds}\frac{dx^\mu}{ds}-\frac{\der^\mu\Omega}{\Omega}.\nonumber\eea Only the null-geodesics are not transformed by the conformal transformations. But, this latter fact also contradicts the claimed geodesic completeness in \cite{bars, bars', bars'', bars-ult}: since the null-geodesics are not transformed by the Weyl gauge transformations, given that these are incomplete in the gauge where a given spacetime sigularity is present, these will be also incomplete in any of the conformally-related gauges. A similar argument signaling the incorrectness of the claim by the authors of \cite{bars, bars', bars''}, but using gauge-independent curvature invariants (in particular the one related with the Weyl invariant $C^2$), is found in \cite{carrasco-kallosh} (see the related discussion in section \ref{sing-sub} of this paper). The above controversy illustrates the confusion which may arise when exploring scale invariance if ignore the unavoidable discussion on the geometrical aspects of the given (metric) theory of gravity.

We have demonstrated that, perhaps, Weyl-integrable geometry -- being the result of lifting Riemann geometry to the category of a Weyl-invariant geometrical theory -- may be a better suited arena where to play the scale-invariant laws of gravity. Several consequences of such a fully Weyl-invariant theory of gravity and of particles for cosmology, as well as for particle's physics, are explored in \cite{quiros-2014}. 

The author thanks David Stefanyszyn for helpful comments and the SNI of Mexico for support of this research.

\end{document}